\documentclass[12pt,twoside]{article}

\usepackage{amsmath}

\usepackage{amssymb}

\usepackage{cite}

\date{}

\begin{document}

\centerline{}

\centerline{}

\centerline {\Large{\bf A necessary and sufficient condition }}

\centerline{}

\centerline{\Large{\bf for a graph $G$, which satisfies }}

\centerline{}

\centerline{\Large{\bf the equality $\mu_{21}(G)=|V(G)|$}}

\centerline{}

\centerline{\bf {Narine N. Davtyan}}

\centerline{}

\centerline{Ijevan Branch of Yerevan State University, Ijevan,
Republic of Armenia}

\centerline{nndavtyan@gmail.com}

\centerline{}

\centerline{\bf {Rafayel R. Kamalian}}

\centerline{}

\centerline{Ijevan Branch of Yerevan State University, Ijevan}

\centerline{The Institute for Informatics and Automation Problems of
NAS RA,}

\centerline{Yerevan, Republic of Armenia, rrkamalian@yahoo.com}

\newcommand{\NDot}{\hspace{-6pt}\textbf{.}\hspace{2pt}}
\newtheorem{Theorem}{Theorem}
\newtheorem{Corollary}{Corollary}
\newtheorem{Lemma}{Lemma}
\newtheorem{Proposition}{Proposition}

\centerline{}

\begin{abstract}
A necessary and sufficient condition is found for a graph $G$, which
satisfies the equality $\mu_{21}(G)=|V(G)|$.
\end{abstract}

\section{Introduction}

We consider undirected, simple, finite and connected graphs, which
contains at least one edge. The terms and concepts which are not
defined can be found in \cite{West1}. The sets of vertices and edges
of a graph $G$ are denoted, respectively, by $V(G)$ and $E(G)$. The
degree of a vertex $x\in V(G)$ is denoted by $d_G(x)$. Let
$\Delta(G)$ be the maximum degree of a vertex of $G$. A function
$\varphi:E(G)\rightarrow \{1,...,t\}$ is called a proper edge
$t$-coloring of a graph $G$, if adjacent edges are colored
differently and each of $t$ colors is used. The least value of $t$,
for which there exists a proper edge $t$-coloring of a graph $G$ is
denoted by $\chi'(G)$. For any graph $G$, and for any integer $t$,
satisfying the inequality $\chi'(G)\leq t\leq|E(G)|$, we denote by
$\alpha(G,t)$ the set of all proper edge $t$-colorings of $G$. Let:
$$
\alpha(G)\equiv\bigcup_{t=\chi'(G)}^{|E(G)|}\alpha(G,t).
$$

An arbitrary nonempty finite subset of consecutive integers is
called an interval. If $\varphi\in\alpha(G)$ and $x\in V(G)$, then
the set of colors of edges incident with $x$ is denoted by
$S_G(x,\varphi)$. If $G$ is a graph, $\varphi\in\alpha(G)$, then
$f_G(\varphi)\equiv|\{x\in V(G)/S_G(x,\varphi) \textrm{ is an
interval}\}|$.

For a graph $G$ and any integer $t$, we define \cite{Mebius6}:
$$
\mu_1(G,t)\equiv\min_{\varphi\in\alpha(G,t)}f_G(\varphi),\qquad
\mu_2(G,t)\equiv\max_{\varphi\in\alpha(G,t)}f_G(\varphi).
$$

For any graph $G$, we set:
$$
\mu_{11}(G)\equiv\min_{\chi'(G)\leq t\leq|E(G)|}\mu_1(G,t),\qquad
\mu_{12}(G)\equiv\max_{\chi'(G)\leq t\leq|E(G)|}\mu_1(G,t),
$$
$$
\mu_{21}(G)\equiv\min_{\chi'(G)\leq t\leq|E(G)|}\mu_2(G,t),\qquad
\mu_{22}(G)\equiv\max_{\chi'(G)\leq t\leq|E(G)|}\mu_2(G,t).
$$

Clearly, these parameters are correctly defined for an arbitrary
graph. Some results on them were obtained in \cite{Simple7, Akunq,
Mebius6, Minchev, Arpine8, Arpine9, Arpine10, Arpine11, Nikolaev12,
Petersen, Evg13, Trees14, Algorithm, Tree_Kontr}.

$\varphi\in\alpha(G)$ is called an interval edge coloring of a graph
$G$ if $f_G(\varphi)=|V(G)|$. The set of all graphs, for which there
exists an interval edge coloring, is denoted by $\mathfrak{N}$. For
any graph $G\in\mathfrak{N}$, we denote by $w(G)$ and $W(G)$,
respectively, the least and the greatest value of $t$, for which $G$
has an interval edge $t$-coloring.

For arbitrary integers $n$ and $i$, satisfying the inequalities
$n\geq3$, $2\leq i\leq n-1$, and for any sequence
$A_{n-2}\equiv(a_1,a_2,\dots,a_{n-2})$ of nonnegative integers, we
define the sets $V[i,A_{n-2}]$ and $E[i,A_{n-2}]$ as follows:
$$
\begin{array}{l}
V[i,A_{n-2}]\equiv\left\{
\begin{array}{ll}
\{y_{i,1},\dots,y_{i,a_{i-1}}\}, & \textrm{if $\;a_{i-1}>0$}\\
\emptyset, & \textrm{if $\;a_{i-1}=0$,}
\end{array}
\right.
\\
E[i,A_{n-2}]\equiv\left\{
\begin{array}{ll}
\{(x_i,y_{i,j}),/\;1\leq j\leq a_{i-1}\}, & \textrm{if $\;a_{i-1}>0$}\\
\emptyset, & \textrm{if $\;a_{i-1}=0$.}
\end{array}
\right.
\end{array}
$$

For any integer $n\geq3$, and for any sequence
$A_{n-2}\equiv(a_1,a_2,\dots,a_{n-2})$ of nonnegative integers, we
define a graph $T[A_{n-2}]$ as follows:
$$
\begin{array}{l}
V(T[A_{n-2}])\equiv\{x_1,\dots,x_n\}\cup\Big(\bigcup_{i=2}^{n-1}V[i,A_{n-2}]\Big),\\
E(T[A_{n-2}])\equiv\{(x_i,x_{i+1})/\;1\leq i\leq
n-1\}\cup\Big(\bigcup_{i=2}^{n-1}E[i,A_{n-2}]\Big).
\end{array}
$$

A graph $G$ is called a galaxy, if either $G\cong K_2$, or there
exist an integer $n\geq3$ and a sequence
$A_{n-2}\equiv(a_1,a_2,\dots,a_{n-2})$ of nonnegative integers, for
which $G\cong T[A_{n-2}]$.

In this paper a necessary and sufficient condition is found for the
equality $\mu_{21}(G)=|V(G)|$.

\section{Preliminary Notes}

\begin{Proposition}\label{p1}\NDot
For an arbitrary galaxy $G$ the following statements hold:
\begin{enumerate}
  \item $G\in\mathfrak{N}$,
  \item $w(G)=\Delta(G)$,
  \item $W(G)=|E(G)|$,
  \item for an arbitrary integer $t$, satisfying the inequality
        $w(G)\leq t\leq W(G)$, there exists
        $\varphi_t\in\alpha(G,t)$ with $f_G(\varphi_t)=|V(G)|$.
\end{enumerate}
\end{Proposition}
\textbf{\textit {Proof}} follows from the results of \cite{Preprint}
(the translation is available on \\
http://arxiv.org/abs/1308.2541v1).

\begin{Proposition}\label{p2}\NDot\cite{Luhansk, Arxiv, MinchevNew}
For any graph $G$, $\varphi\in\alpha(G,|E(G)|)$ with
$f_G(\varphi)=|V(G)|$ exists iff $G$ is a galaxy.
\end{Proposition}

\section{Main Result}

\begin{Proposition}\label{p3}\NDot
For arbitrary galaxy $G$, the equality $\mu_{21}(G)=|V(G)|$ is true.
\end{Proposition}

\textbf{\textit {Proof.}} Let $G$ be an arbitrary galaxy. From
proposition \ref{p1} it follows that for $\forall
t\in[\Delta(G),|E(G)|]$, there exists $\varphi_t\in\alpha(G,t)$ with
$f_G(\varphi_t)=|V(G)|$. It, particularly, means, that for $\forall
t\in[\Delta(G),|E(G)|]$ $\mu_2(G,t)=|V(G)|$. It, by definition,
means that $\mu_{21}(G)=|V(G)|$.

\textbf{\textit {The proposition is proved.}}

\begin{Theorem}\label{t1}\NDot
For any graph $G$, the equality $\mu_{21}(G)=|V(G)|$ holds if and
only if $G$ is a galaxy.
\end{Theorem}

\textbf{\textit {Proof.}} Let $G$ be a galaxy. From proposition
\ref{p3} it follows that $\mu_{21}(G)=|V(G)|$.

Let $G$ be a graph with $\mu_{21}(G)=|V(G)|$. Let us show that $G$
is a galaxy. Assume the contrary: $G$ is not a galaxy. From
proposition \ref{p2} we obtain that for
$\forall\varphi\in\alpha(G,|E(G)|)$ $f_G(\varphi)\leq|V(G)|-1$. It
means that $\mu_2(G,|E(G)|)\leq|V(G)|-1$. Now, taking into account
the definition of the parameter $\mu_{21}$, we conclude that
$\mu_{21}(G)\leq|V(G)|-1$. Contradiction.

\textbf{\textit {The theorem is proved.}}

\end{document}